\documentclass[conference, letterpaper]{IEEEtran}

\newcommand{\dashboard}[1]{\mbox{DAS-Dashboard}}
\newcommand{\gateway}[1]{GW#1}
\newcommand{\dataServer}[1]{Data Analytics Server}

\newcommand{\AcceptedThreshold}[1]{\ensuremath{\tau}#1}
\newcommand{\LearningRate}[1]{\ensuremath{\alpha}#1}
\newcommand{\DiscountFactor}[1]{\ensuremath{\gamma}#1}

\usepackage{amsmath}
\usepackage{mathtools}
\usepackage{dsfont}
\usepackage{paralist}
\usepackage[table]{xcolor}
\usepackage{footnote}
\usepackage{array}
\usepackage{ragged2e}
\usepackage{bm}
\usepackage{textcomp}
\usepackage{hyperref}

\usepackage{subcaption}
\usepackage{mathtools}
\usepackage{algorithmic}

\usepackage{verbatim}

\usepackage[acronym]{glossaries}
\newacronym{wsn}{WSN}{Wireless Sensor Network}

\ifCLASSINFOpdf
  \usepackage[pdftex]{graphicx}
  \graphicspath{{figures/}{plots/}}
\else
\fi

\usepackage{rotating}
\usepackage{multirow}

\begin{document}


%
\title{Cloud Empowered Self-Managing WSNs}

\author{
\IEEEauthorblockN{Gabriel Martins Dias\IEEEauthorrefmark{1},
    Cintia Borges Margi\IEEEauthorrefmark{2},
    Filipe C. P. de Oliveira\IEEEauthorrefmark{2} and
    Boris Bellalta\IEEEauthorrefmark{1}}
\and\and
\IEEEauthorblockA{
    \hspace{1cm}\IEEEauthorrefmark{1}Department of Information and Communication Technologies\\
    \hspace{1cm} Universitat Pompeu Fabra\\
    \hspace{1cm}Barcelona, Spain\\
    \hspace{1cm}\{gabriel.martins, boris.bellalta\}@upf.edu}
\and
\IEEEauthorblockA{
    \IEEEauthorrefmark{2}Escola Politecnica \\
		Universidade de Sao Paulo \\
		Sao Paulo, Brazil\\
    \{cintia,  filipe.calasans.oliveira\}@usp.br}
}

\maketitle

\begin{abstract}
Wireless Sensor Networks (WSNs) are composed of low powered and resource-constrained wireless sensor nodes that are not capable of performing high-complexity algorithms. Integrating these networks into the Internet of Things (IoT) facilitates their real-time optimization based on remote data visualization and analysis. This work describes the design and implementation of a scalable system architecture that integrates WSNs and cloud services to work autonomously in an IoT environment. The implementation relies on Software Defined Networking features to simplify the WSN management and exploits data analytics tools to execute a reinforcement learning algorithm that takes decisions based on the environment’s evolution. It can automatically configure wireless sensor nodes to measure and transmit the temperature only at periods when the environment changes more often. Without any human intervention, the system could reduce nearly $85\%$ the number of transmissions, showing the potential of this mechanism to extend WSNs lifetime without compromising the data quality. Besides attending to similar use cases, such a WSN autonomic management could promote a new business model to offer sensing tasks as a service, which is also introduced in this work.


\end{abstract}


\section{Introduction}

Wireless sensor nodes are power supplied by batteries and equipped with radio
antennas and sensors that are capable of sensing several environmental 
parameters.
Thanks to their portable size and sensing capabilities, sensor nodes are often 
densely deployed in areas that are not necessarily humanly accessible.
After deployed, sensor nodes start to collect and report environmental data, 
which explains why Wireless Sensor Networks (WSNs) are considered data-oriented 
networks: sensed data is the most valuable asset that monitoring WSNs can 
produce.

Fine-tuning a WSN requires appropriate management to detect and handle 
several types of problems.
For instance, if a sensor node runs out of battery, a new routing table must be 
generated, changing the topology and affecting other sensor nodes in the 
vicinity.
Alternatively, if a sensor node's radio is experiencing a temporary interference, 
its transmissions will suffer sporadic errors, but the routing table 
may be maintained to avoid the overhead communication that would provoke 
extra transmissions and probably worsen the data delivery.


Similarly to WSNs, Internet of Things (IoT) environments are usually composed 
by smart devices that communicate between themselves, and part of their 
communication may occur to transmit environmental parameters from sensing 
devices, such as temperature, relative humidity and solar radiation.
An IoT environment is characterized by its capacity of interconnecting 
"things", which can be represented by household appliances, machines, 
personal devices or living beings.
Such environments do not require human intervention to work properly, 
because they are able to self-configure and self-manage resources in reaction 
to external phenomena that impact their operation.
Furthermore, IoT applications have a broader scope than traditional WSNs: 
devices can be more powerful and compute high complexity algorithms, interact with 
humans, provide machine-to-machine communication and also connect to cloud 
services to extend their computing power.
In conclusion, rather than simply using smart devices as sensors that 
periodically report measurements, WSNs can be incorporated to IoT environments 
and, consequently, empowered by cloud services~\cite{Bellavista2013}.

Among several cloud services, data analysis can be ranked as the most 
relevant for WSNs, given the sensor nodes' dedication to measure and report 
information from the external world.
The ease of access to data analysis services is a twofold facility,
because analyzing data in runtime does not only provide information about the
monitored environment, but it also facilitates the evaluation of the quality of
the service provided by sensor networks.

However, each WSN has its own requirements and particularities, such as maximum 
delay to deliver the data, tolerance about node failures and unreliable 
transmissions.
Thus, managing multiple WSNs in a centralized architecture would require 
several configurations and parallel tasks that could overload and collapse a 
server that must handle dozens (occasionally hundreds) of sensor nodes at the 
same time.
Moreover, as data cannot be properly handled in WSNs (due to the constrained 
sensor nodes' computing power and limited access to external resources), a proper 
data analysis manager must be able to manipulate several datasets simultaneously, 
which requires high memory and processing capabilities.
Hence, a scalable solution that integrates WSNs in IoT environments must 
lean on specialized knowledge about the advantages and disadvantages in 
fine-tuning WSNs and data handling.

This work describes an architecture that integrates solutions in different 
planes to incorporate WSNs in IoT environments.
The first component of such an architecture is a WSN application and resource 
manager, which is responsible for managing WSNs' resources and abstracting 
sensor nodes in an application layer accessible for any system or person 
in charge of managing the WSNs.
In practice, sensor nodes' detailed operation remains transparent to
any other system external to the WSNs.

In the proposed architecture, the information that would be provided to WSN 
managers is actually communicated to a central dashboard that provides means 
for collecting, storing and publishing data transmitted by wireless sensor 
nodes.
The dashboard can delegate the WSN management and the data analysis to 
appropriate mechanisms that may perform their operations remotely in the cloud, 
keeping their customization to meet WSN applications' requirements.
Finally, a \dataServer{} supports different data analytics tools and 
algorithms that can optimize the data collection in terms of Quality of 
Information.


\begin{figure}[t]
        \centering
	\includegraphics[width=0.45\textwidth]{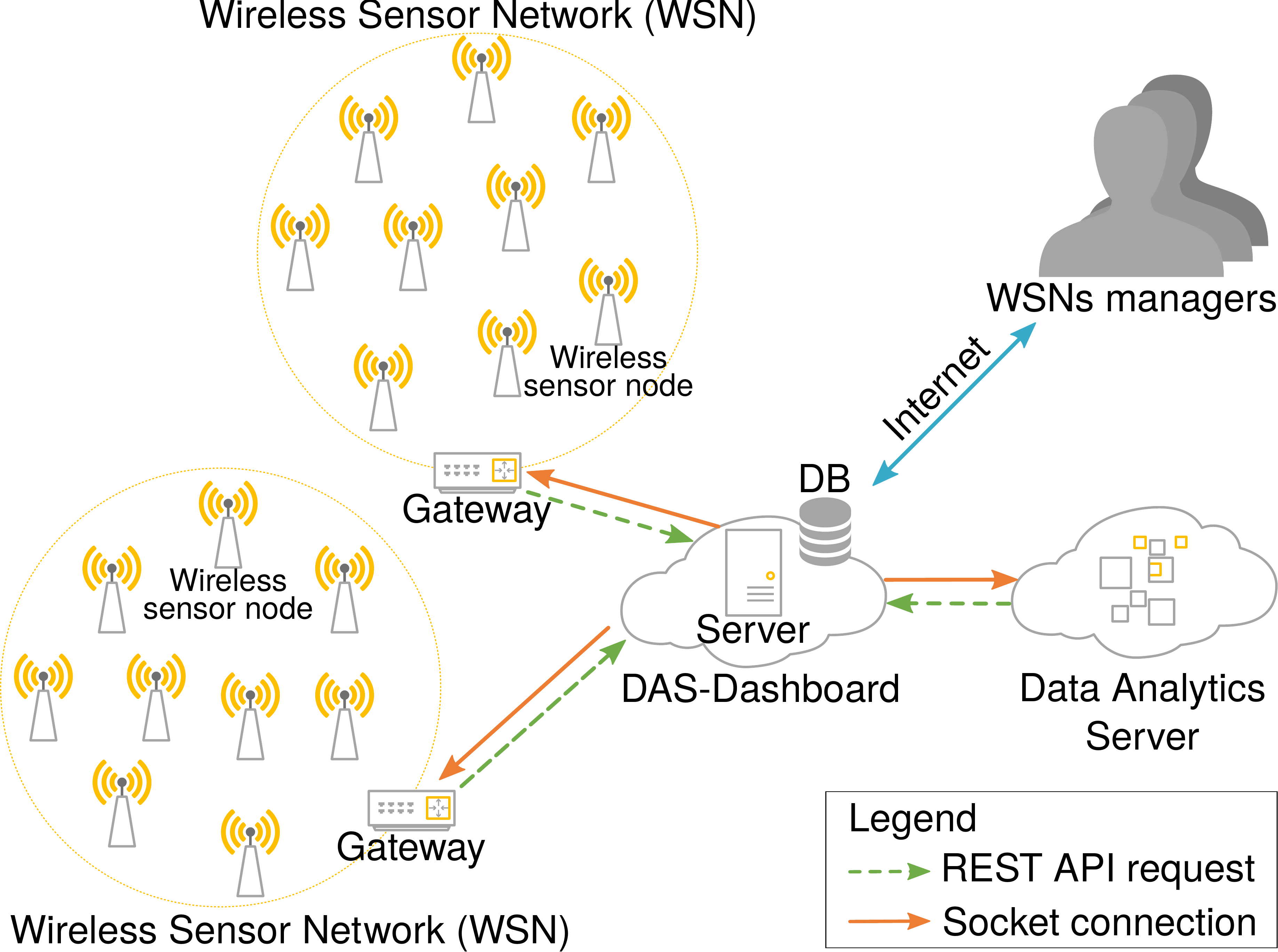}
        \caption{The self-managing architecture for WSNs.}
        \label{fig:system-architecture}
\end{figure}

Supported by TCP connections established among different components, each 
component of this architecture can be moved to the cloud and become
remotely reachable.
Such an accessibility permits WSNs to interact with other WSNs, communicate 
with other systems and simultaneously serve several users.
Moreover, users do not need to spend time to adjust WSNs with the 
best set of parameters, to configure protocols to enhance their power savings 
or to \mbox{(re-)synchronize} network components after sensor nodes 
replacement, because all these tasks are automatically--and 
properly--guaranteed by the WSN resource managers.

The proper integration of WSN resource managers and data analytics tools 
results on a self-managing architecture that controls WSNs based on real time 
data analysis, as shown in Figure~\ref{fig:system-architecture}.







\begin{figure*}[t]
 \centering
\begin{subfigure}[t]{0.21\textwidth}        
	\includegraphics[width=\textwidth]{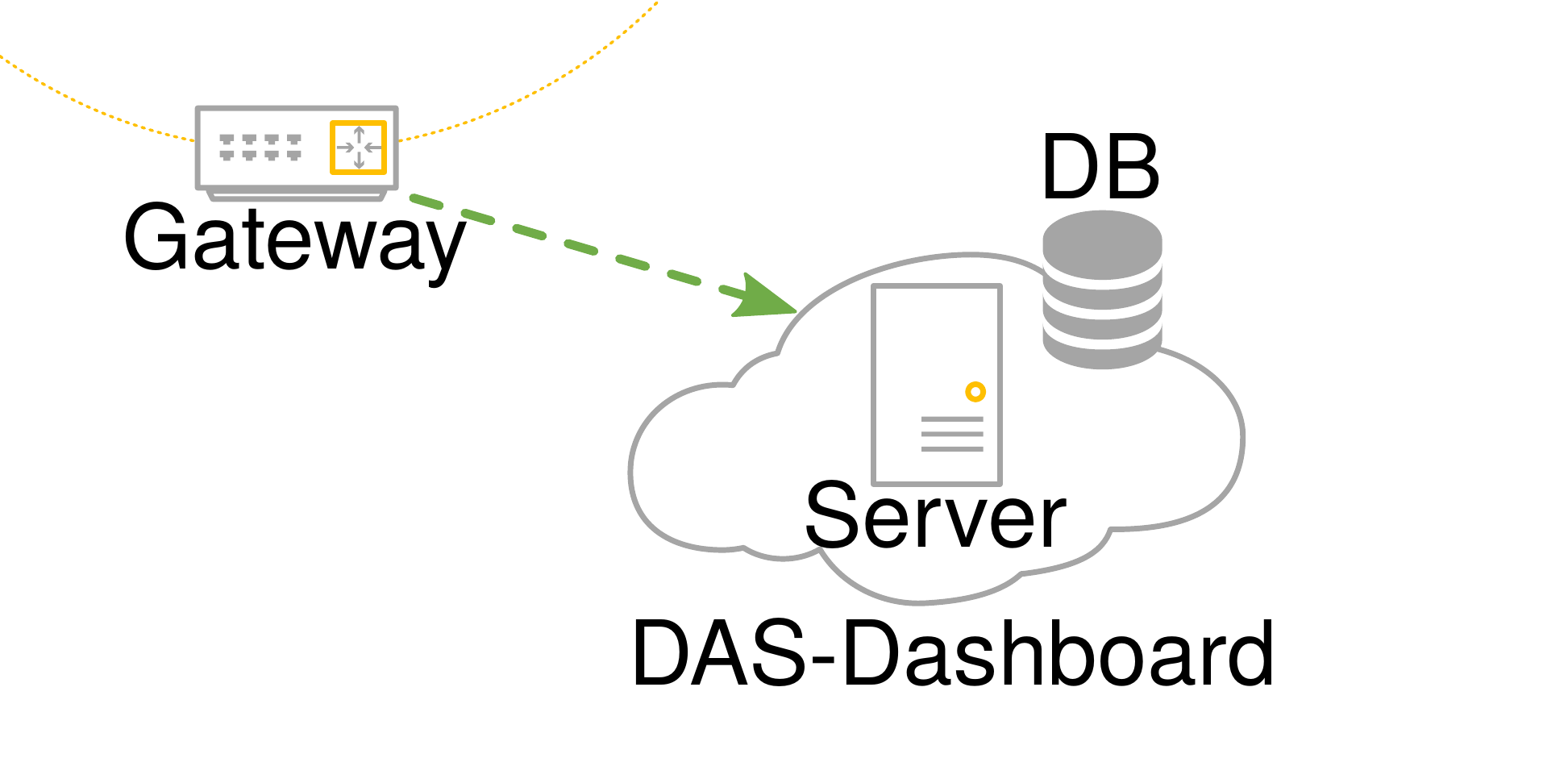}
        \caption{\textbf{Step 1}. Sensors perform their default sensing tasks, 
before transmitting their measurements via radio to \gateway{s}. Data is 
reported to the \dashboard{} and stored in the database for further access.}
        \label{fig:loop-monitor}
\end{subfigure}
 \qquad
\begin{subfigure}[t]{0.22\textwidth}
	\includegraphics[width=\textwidth]{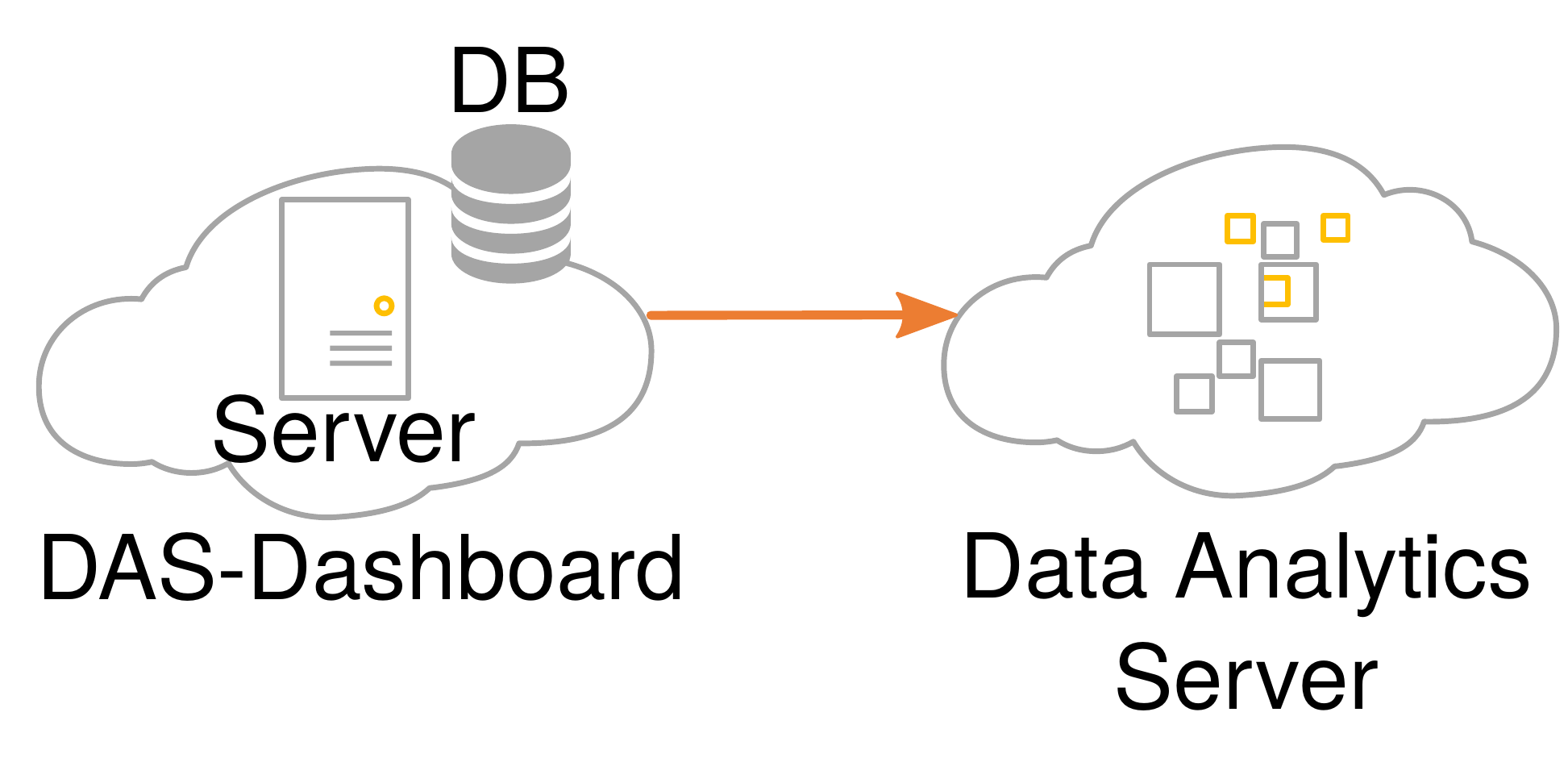}
        \caption{\textbf{Step 2}. Measurements received by the \dashboard{} 
are communicated to the \dataServer{}. 
The \dataServer{} perform the data analysis to infer whether a sensor is 
gathering 
informative data or not.}
        \label{fig:loop-analysis}
\end{subfigure}
\qquad
\begin{subfigure}[t]{0.22\textwidth}
	\includegraphics[width=\textwidth]{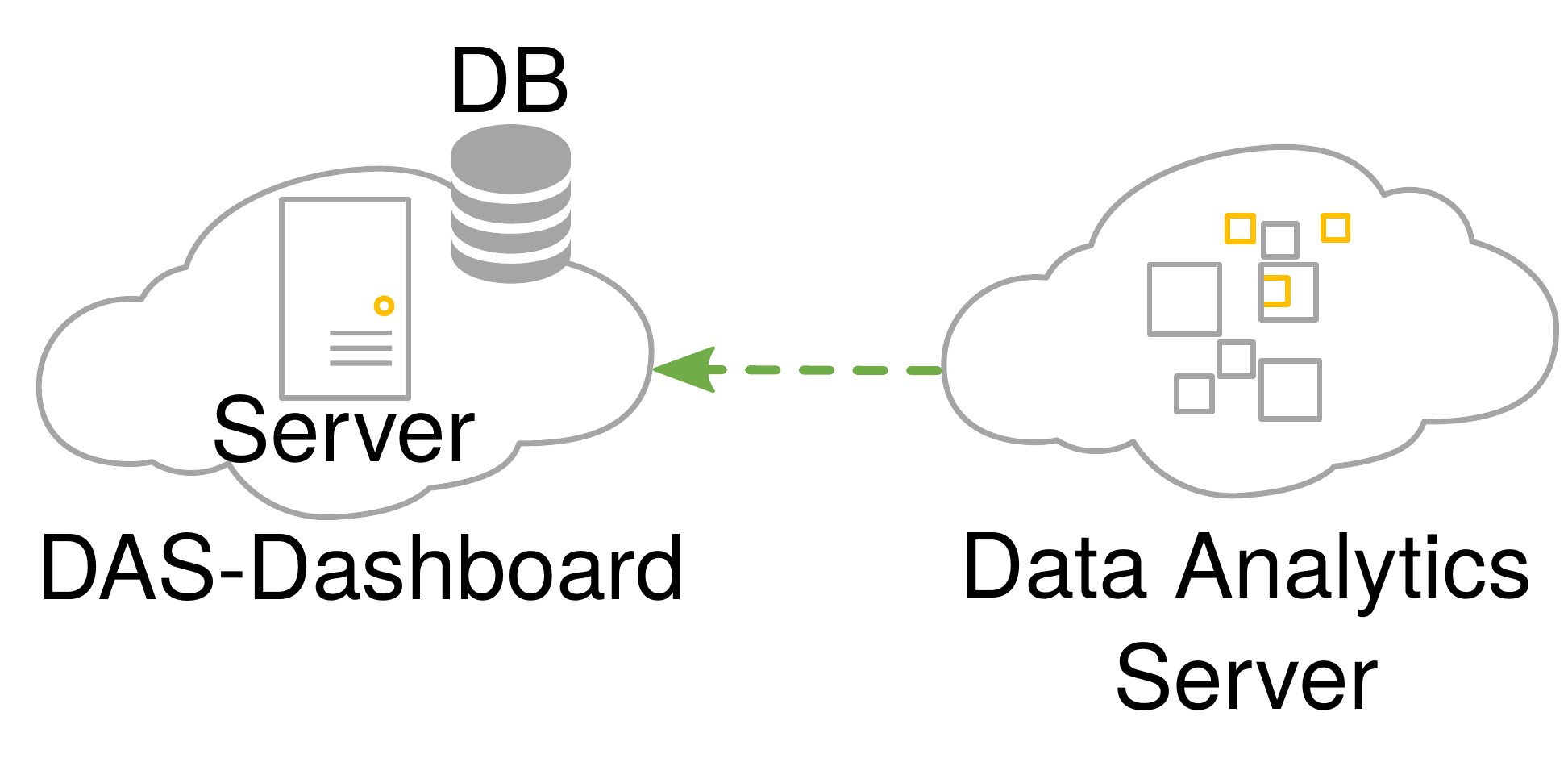}
        \caption{\textbf{Step 3}. The \dataServer{} transmit to the 
\dashboard{} new plans for the WSN, depending on the application type. These 
plans may be linked with network details (such as sensor nodes' positions) and 
can rely on external factors (such as the time of the day).}
        \label{fig:loop-plan}
\end{subfigure}
\qquad
\begin{subfigure}[t]{0.21\textwidth}
	\includegraphics[width=\textwidth]{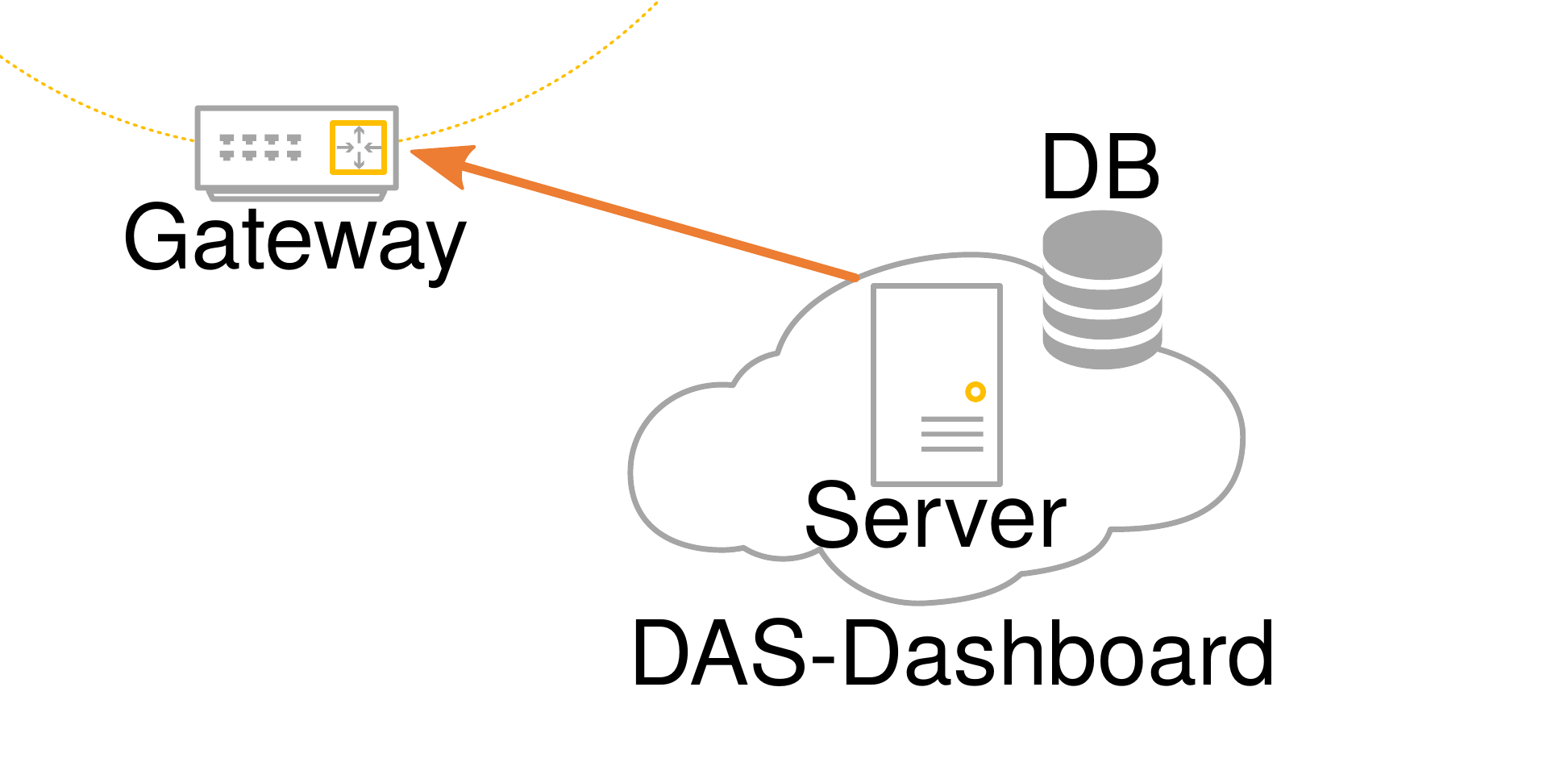}
        \caption{\textbf{Step 4}. At this point, the \dataServer{} announces 
the recommendation to the \dashboard{}. 
Sensor nodes are reprogrammed according to the the instructions communicated 
to their \gateway{s}.}
        \label{fig:loop-execute}
\end{subfigure}
\caption{The self-managing architecture in detail. }
\label{figure:loop}
\end{figure*}

\section{Architectures for Scalable WSNs}



IoT environments have an information flow that extends the
standard WSN data collection, because WSNs' operation can be improved 
based on data analysis executed in runtime.
The WISEBED project has already addressed the need of a common
platform for programming WSNs and collecting data from sensor 
nodes~\cite{Chatzigiannakis2010}.
In that project, several testbeds were deployed, and remote users were able
to program sensor nodes and run experiments to prove concepts 
previously evaluated only in simulations.
This solution, however, did not integrate tools for remote data analysis 
and real-time WSN optimization.

More recently, a new architecture was proposed to integrate WSNs in IoT
environments~\cite{Ezdiani2016}.
There, a central server communicates with sensor nodes and receive
their statistics, which are analyzed by a third-party tool every hour.
Based on the data analysis, the central server is able to react to 
dynamic changes in network conditions and provide flexibility to the
WSN, through remote reconfiguration of the sensor nodes.
As a drawback, their architecture is not scalable, because the management 
in the central server is highly coupled.
In summary, the server must have installed the same Operating System (OS) 
used in the sensor nodes, besides hosting the database to store collected 
data and the tool used for data analysis.
Hence, the system's growth is bounded by the capacity of a unique central 
server to scale up and integrate many WSNs, data analytics tools or give 
access to several remote users simultaneously.

The architecture proposed in this work aims for scalability.
It relies on the power of shifting most of the computation to the cloud, as 
detailed in Figure~\ref{figure:loop}.
First, sensor nodes must report data to their respective Gateways 
(\gateway{s}), which is connected to the central server via Internet.
The data provided by sensor nodes can be statistics about network conditions,
such as delays, packet losses and data throughput, or their measurements.
After analyzing the reported values, 
the \dataServer{} may generate a report and recommend the most proper
changes to the WSN operation.
Finally, the new WSN's operation plan is executed after sensor nodes are 
updated to adjust their tasks according to the results of the data analysis.

To test an use case that may benefit from data analysis, a WSN was 
deployed to measure temperature in an office, where values vary less often 
during the night. Therefore, it is not necessary to sample as often as during 
the day, when the air conditioning system and the presence of people impact the 
environment.



\subsection{WSN Application and Resource Manager}

Wireless sensor nodes are typically close to the data origin and have 
constrained computing capabilities to store and process information.
In homogeneous WSNs with similar wireless sensor nodes (in terms of software 
and hardware), each sensor node may have different configurations, according to 
its localization and measurements' relevance.
In heterogeneous WSNs, sensor nodes may differ in more ways, such as computing 
capabilities, OSs and clustering roles.
Because of these particularities, addressing the architecture's scalability and 
aiming for simultaneously controlling multiple WSNs requires a framework that 
can avoid resource underutilization and handle eventual changes in WSNs' 
topology.
These aspects facilitate sensor nodes (re)placement and favor the expansion 
and evolution of WSNs in terms of hardware and software.

For example, TinyDB~\cite{Madden:2005:TAQ:1061318.1061322} 
is a framework that allows sensor nodes to be queried to periodically measure 
environmental parameters, such as temperature and humidity.
That is, given a set of queries, TinyDB is able to analyze and optimize the use 
of the WSN's resources, shortening delivery routes and reducing the overall 
energy consumption.
However, even though TinyDB considers the sensor nodes' energy consumption 
to decide for the most suitable execution plan, it does not adapt to topology 
changes that may also impact the routing and the end-to-end 
delays in the data delivering, which does not favor the scalability of the 
WSNs.

Similarly to TinyDB, the DIstributed Self-Organizing NEtwork management 
(DISON)~\cite{Minh2013} is a framework that permits WSNs to 
self-organize, react to changes in the topology and optimize the 
overall energy consumption.
However, DISON forces each sensor node to reconfigure its operations according 
to its own resources and the network state, which may become an
overhead in larger WSNs.

\begin{figure}[t]
        \centering
	\includegraphics[width=0.45\textwidth]{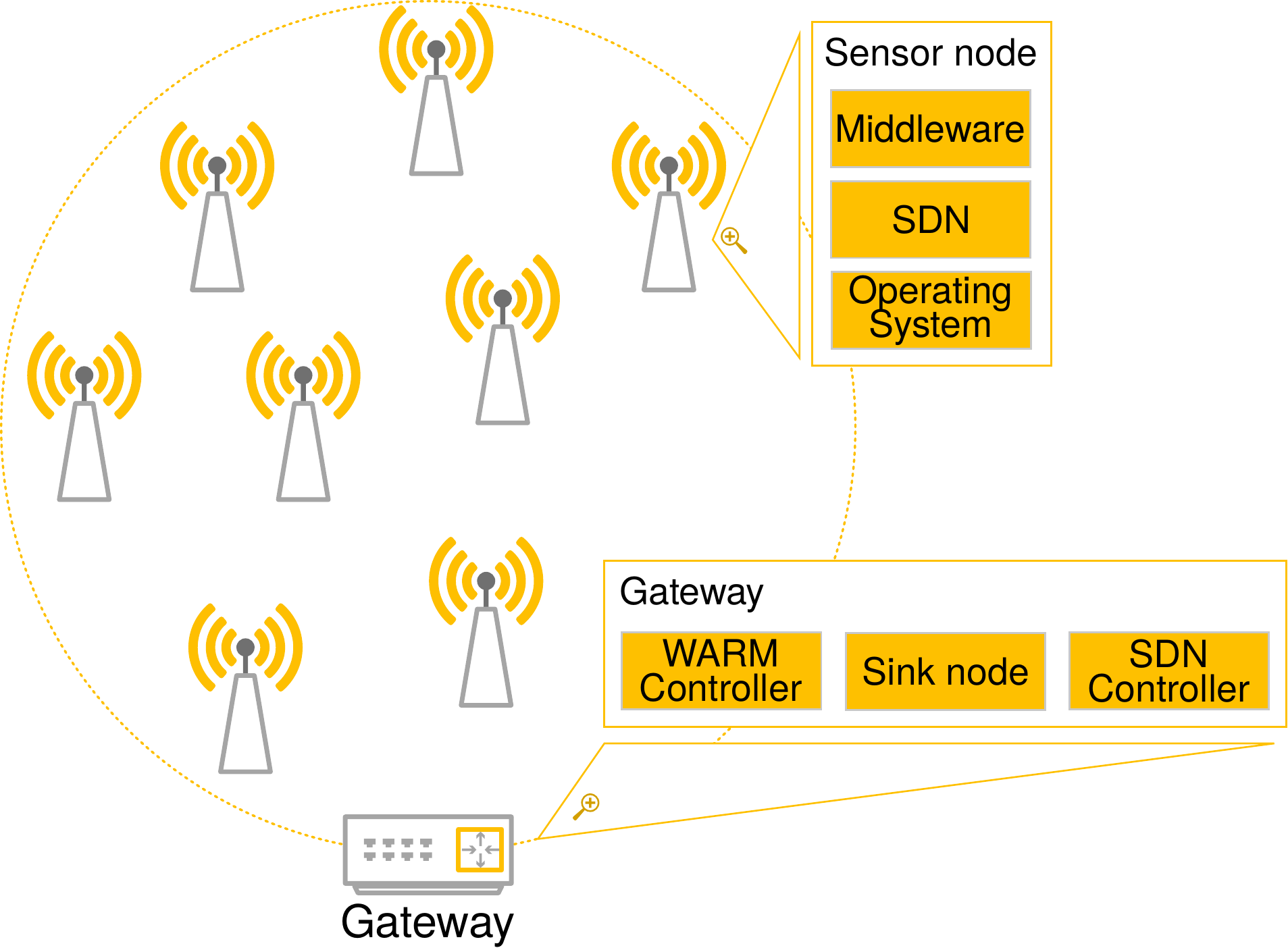}
        \caption{The WARM architecture disposed in the WSN.}
        \label{fig:warm-architecture}
\end{figure}

The architecture proposed in this work is focused on supporting several sensor 
node types and tasks, while reducing the management control at the most.
The scalability of this system can be provided by a framework that abstracts 
sensor nodes at the data plane and reduces management tasks, such as
the WSN Application development and Resource Management 
(WARM)~\cite{Silv1608:WARM}.
WARM relies on Software Defined Networking (SDN) features to simplify the 
development and resource management for WSN applications.
The SDN controller implemented in sensor nodes abstracts the control and data 
layers~\cite{DeOliveira2014}, which facilitates the adoption of low level tasks 
as high-level network applications that can be easily configured.
Moreover, this separation keeps the network control in the SDN controller, 
which is responsible for dealing with topology changes and exploring their 
resources at the best.
To meet the scalability requirements described above, WARM was adopted as the 
WSN application and resource manager.
Figure~\ref{fig:warm-architecture} shows the WARM's architecture in detail.

To control the WSN, a sensor node with the WARM controller installed is 
responsible for receiving via serial port any configuration and management 
commands from the WSN manager.
These commands are further translated to the format used by the SDN controller.
Besides transmitting commands, the WARM controller communicates eventual 
notifications from sensor nodes to the WSN manager, such as new sensor nodes 
associations and acknowledgments after updates.

To configure and manage WSN applications, WARM provides a hardware-independent 
application layer, which makes wireless sensor nodes' particularities 
transparent to any component external to the WSN.
Meanwhile, sensor nodes can still be controlled via an interface that provides 
means to retrieve their status and schedule tasks to periodically sense 
environmental parameters or compute collected data.
For example, sensor nodes can be programmed to execute lightweight processing 
algorithms, environmental sensing or actuator actions, according to their 
individual capabilities.
%
In practice, to schedule a new sensing task, the WSN manager needs only to 
inform which node will receive the measurements and the time between 
consecutive measurements.
WARM also specifies interfaces to support new applications that can be further 
designed by WSN managers.


\subsection{\dashboard{}}

WSN managers may adopt different strategies to monitor and analyze all the 
information retrieved by their WSNs.
Storing collected data in databases allows further access to historical 
information, besides providing data visualization to other systems.
If the database is connected to the Internet, collected data can be available 
to remote users, overcoming physical limitations and providing access to 
external systems that may independently process such information.

As a drawback, setting up a new database for each WSN implies overhead, because 
their schema must be designed, a server must be configured and the 
communication with the WSN must be properly established.
A standard storage medium can make the data handling transparent to the WSN
manager, removing the overhead to customize data formats and facilitating the
communication with other systems, which allows WSN managers to outsource 
data processing, such as filtering and analyzing the collected data, and, 
especially, predicting future measurements.

To communicate with WSNs to collect, store and publish all data transmitted by 
wireless sensor nodes, the \emph{Data Analytics for Sensors 
Dashboard} (\dashboard{})~\cite{Dias2016c} was chosen.
The \dashboard{} can receive two types of input: \begin{inparaenum}[(i)] 
\item values reported by sensor nodes; and 
\item recommendations generated by an external prediction system.
\end{inparaenum}
To store the data, a PostgreSQL database server was connected to the system's 
back-end, allowing further access to historical information.
The data input is handled by the back-end, implemented in 
Sails.js\footnote{\href{http://sailsjs.org}{http://sailsjs.org}},
which facilitates the creation of APIs to manage the insertion of 
new information and to provide further on-demand access via HTTP Requests.
The use of APIs allows the communication among different servers 
and also guarantees the loose coupling with the front-end.
The front-end, implemented in 
AngularJS\footnote{\href{https://angularjs.org}{https://angularjs.org}}, 
provides data visualization to network managers and remote users via 
Internet, which is not possible in other IoT platforms, such 
as the Realtime.io\footnote{\href{http://realtime.io}{http://realtime.io}} and 
the nimbits Platform\footnote{\href{http://bsautner.github.io/com.nimbits}{
http://bsautner.github.io/com.nimbits}}.

The \dashboard{} guarantees that the data inserted in the database is 
communicated to other servers in the form of events via socket connections to 
services (registered in advance) that keep listening for changes during the WSNs 
lifetime.
Establishing standards for data insertion (API requests) and data publishing 
(socket connections) allows the \dashboard{} to integrate with Big Data services 
and update the operation of the sensor nodes according to the data that they 
have reported.

\subsection{\dataServer{}}

As explained before, typical wireless sensor nodes have constrained memory and 
computing power, besides limited communication with external networks.
Therefore, they do not have enough capacity to store measurements or 
execute high-complexity algorithms to analyze the environment's evolution and 
properly optimize their data collection.
For example, a sensor node could use its own data to forecast if it is going to 
rain and therefore report more often to detail an abrupt change in temperature 
that would occasionally happen.

Specialized tools, such as the Riverbed Modeler~\cite{riverbedmodeler2016}, 
can communicate with the \dashboard{} and receive network statistics, such as 
end-to-end delays, packet arrival times and transmission times.
Later, an optimization plan may be generated, based on simulation results 
obtained in parallel to the real WSN operation.
Alternatively, as the \dashboard{} provides sufficient tools to store and publish 
reported data,  it could be possible to perform data analysis using public 
APIs, such as Google Prediction 
API\footnote{\href{https://cloud.google.com/prediction}
{https://cloud.google.com/prediction}} and Amazon Machine Learning prediction 
APIs\footnote{\href{https://aws.amazon.com/machine-learning}
{https://aws.amazon.com/machine-learning}}.

In this work, a customized \dataServer{} was implemented in 
R\footnote{\href{https://www.r-project.org}{https://www.r-project.org}}.
It can perform different types of data analysis and recommendations, 
occasionally relying on external data resources, such as public services and 
other databases via Internet.
Based on the data analysis results, it generates customized recommendations for 
the sensor nodes, for example, to change the time interval between two 
measurements based on predictions about future measurements.

\section{Experimental Setup}

To test the proposed architecture, a WSN was deployed to monitor 
room-temperature in an office, which is constantly changing.
In this scenario, there are several factors that impact the measurements, 
such as the presence of people and the air conditioning system.
TelosB motes were used in the WSN deployment, and the \dashboard{} has been 
deployed in a well-dimensioned machine without energy nor performance 
constraints, with reliable Internet connection and direct access to the 
WARM controller.

In order to analyze the architecture's self-managing capability,
a Reinforcement Learning (RL) algorithm called Q-Learning~\cite{Watkins1992} 
was adopted to analyze the collected data and adjust sensor nodes' sampling 
intervals according to the environmental behavior in the past.
In summary, if a sensor node is constantly reporting similar values, the 
algorithm may recommend an increase in the period between consecutive 
measurements and avoid unnecessary transmissions.
In this case, sensor nodes would save their battery, which can potentially 
extend the WSN lifetime without affecting the quality of the data stored in the 
\dashboard{}.
Preliminary simulation results showed that nearly $73\%$ of the transmissions 
could be avoided without compromising the quality 
of the information provided to the WSN manager~\cite{Dias2016d}.

\begin{figure}[t]
        \centering
	\includegraphics[width=0.45\textwidth]{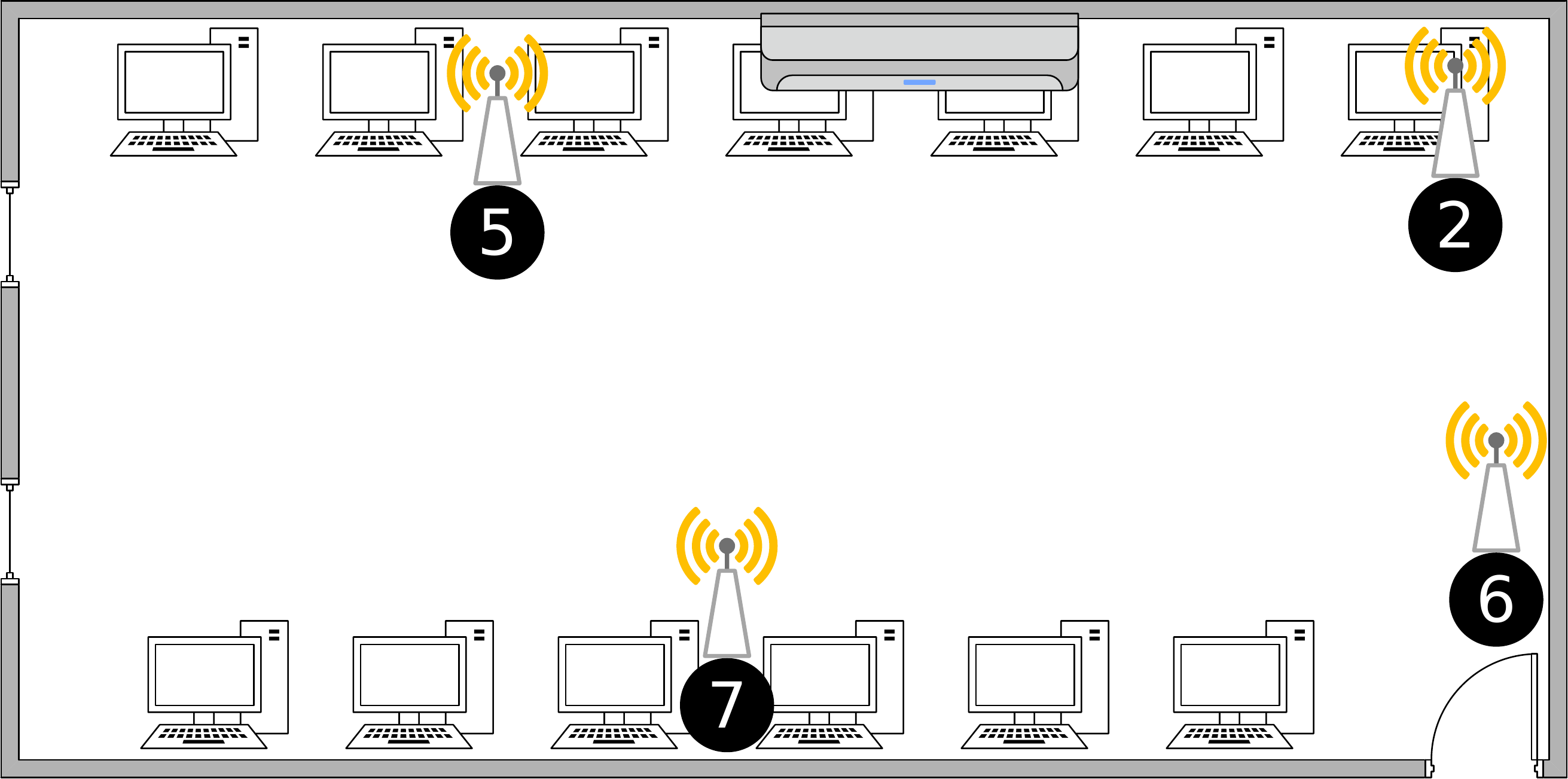}
        \caption{Sensor nodes' position in the office}
        \label{fig:room}
\end{figure}

During $2$ days, the room-temperature data was collected using $4$ wireless 
sensor nodes placed in the office, as shown in Figure~\ref{fig:room}.
After that, the \dataServer{} was activated to systematically setup the 
most proper sampling interval so as to guarantee the best quality-resource 
trade-off under the current environmental conditions.
As for the quality, we set the goal of the agent in terms of an 
\textit{acceptance threshold} \AcceptedThreshold{}, which represents the maximum 
absolute difference between two measurements ($\delta$) tolerated by the 
algorithm.
Based on the data collected in the first days, \AcceptedThreshold{} was set to 
$0.5^\text{o}C$.

Considering measurements from all nodes during the first $2$ days, around 
$6.4\%$ of consecutive measurements made in intervals of $480$ 
seconds would differ by more than this value, and around $0.15\%$ of 
consecutive measurements made in intervals of $30$ seconds would still 
differ by more than this value.
Therefore, we do not expect to measure sufficiently frequent to have all 
measurements differ by less than \AcceptedThreshold{}.
However, we know that there are several cases in which sensors do not have to 
sample every $30$ seconds, because the measurements would be very similar
in most of the time.
We highlight that higher sampling intervals are preferred to reduce the number 
of transmissions and, consequently, the energy consumption in the sensor nodes.

In this experiment, the configuration parameters of the Q-Learning algorithm 
based on the best results obtained in previous simulations: 
\emph{learning rate} $\LearningRate{} = 0.9$ and \emph{discount factor} 
$\DiscountFactor{} = 0.1$.
To illustrate the results, the first $12$ hours were considered as the time 
necessary for the Q-Learning algorithm to ``calibrate'' the action-value 
function, i.e., when it visits all possible action-state pairs to find the best 
actions that should be taken in the future.
The experiment ran for $30$ hours uninterruptedly.

%

\subsection{Transmissions Savings}


In this scenario, the number of transmissions in a sensor node achieves 
its maximum when the sampling interval is $30$ seconds and minimum when the 
sampling interval is $480$ seconds.
Intuitively, setting the sampling interval to $60$, $120$, $240$ and $480$ 
seconds will represent a reduction of respectively $50\%$, $75\%$, $87.5\%$ 
and $93.75\%$ in the maximum number of transmissions.

\begin{table}[t]
\centering
\def\arraystretch{1.2}
\rowcolors{2}{white}{gray!25}
\begin{tabular}{
	>{\centering\arraybackslash}m{1cm}
	>{\centering\arraybackslash}m{2cm}
	>{\centering\arraybackslash}m{2cm}
	>{\centering\arraybackslash}m{2cm}
	}
  \cellcolor{gray!50}\textbf{Sensor node} &
  \cellcolor{gray!50}\textbf{Transmissions saved} &
  \cellcolor{gray!50}\textbf{$\delta > \AcceptedThreshold{}$}&
  \cellcolor{gray!50}\textbf{Average $\delta$}\\ 
$2$ & $82.71\%$ & $7.68\%$ & $0.29$ \\ 
  $5$ & $84.62\%$ & $4.52\%$ & $0.25$ \\ 
  $6$ & $75.82\%$ & $11.32\%$ & $0.29$ \\ 
  $7$ & $43.35\%$ & $2.78\%$ & $0.09$ \\ 
   \hline
\end{tabular}
\caption{Transmission savings observed in the experiments}
\label{tab:diff}
\end{table}

Every time that the \dashboard{} reprograms a sensor node, at least one extra 
transmission is made (due to packet losses, commands are retransmitted after a 
delay to guarantee their delivery).
Considering also these transmissions, Table~\ref{tab:diff} illustrates how many 
transmissions could be saved in each case when the Q-Learning algorithm was 
adopted to adjust the sensor nodes' sampling interval.
The baseline is the scenario with the sensor nodes sampling always every $30$ 
seconds without extra transmissions from the \dashboard{}.
In the best case (sensor node $\#5$), the number of transmissions was reduced 
by $84.62\%$ of its maximum.
Moreover, it is possible to see that the average absolute difference between 
consecutive measurements is always smaller than \AcceptedThreshold{} and that 
the number of consecutive measurements that differ by more than 
\AcceptedThreshold{} can be lower than $3\%$.

\subsection{Flow Overhead}

The proposed architecture adds some delay inherent to the communication 
between WARM, the \dashboard{} and the \dataServer{}.
Moreover, there are delays due to data processing in the \dashboard{} and in 
the \dataServer{}, besides the time needed to reprogram the sensor nodes and 
adjust their sampling intervals.
To measure these delays, the \gateway{}, the \dashboard{} and the 
\dataServer{} were deployed in the same machine, which eliminates any 
network delays that could impact their operation.

On average, the process of receiving data in the \dashboard{}, analyzing its 
content in the \dataServer{}, reporting a recommendation back to the 
\dashboard{} and reprogramming the sensor node does not take longer than 
$1.2$ seconds.
Table~\ref{tab:delay} shows the average delays and their respective 
standard deviation ($\sigma$), which can be considered lower bounds for future 
distributed deployments.
Note that when sensor nodes are programmed to sample in fixed time intervals, 
the period between consecutive measurements may vary, due to their clock drift.
In these experiments, sensor nodes sample, on average, nearly $1$ second 
earlier than they should report.
This value is not taken into account when calculating the delays caused by the 
new data flow overhead.

\section{Other Applications}

As mentioned before, WSNs are mainly data-oriented networks, i.e., the 
sensed data is their most valuable asset.
Therefore, it is common to rely on analysis over the collected information to 
take further actions.
There is a wide number of applications that can use this architecture to 
optimize existing WSNs and also to enable new business models.
These applications can involve different algorithms for data analysis, new 
modules that can be attached to the \dashboard{} or new applications that can 
be implemented in WARM.
Examples of future applications that could be deployed using this architecture 
are highlighted in the following.

\subsection{Multi-scope Integration}

Data analysis algorithms can be enhanced with artificial intelligence mechanisms 
that evaluate historical trends and trigger actions, activate machines or 
communicate with other systems according to pre-defined policies.
For example, recommendations generated through data analysis can be targeted to 
WSN managers that would take manual actions, such as adding new sensor nodes or 
replacing existing ones.

\begin{table}[t]
\centering
\def\arraystretch{1.2}
\rowcolors{2}{white}{gray!25}
\begin{tabular}{
	>{\centering\arraybackslash}m{1cm}
	>{\centering\arraybackslash}m{1.4cm}
	>{\centering\arraybackslash}m{1.4cm}
	>{\centering\arraybackslash}m{1.4cm}
	>{\centering\arraybackslash}m{1.4cm}
	}
  \cellcolor{gray!50}\textbf{Sensor node} &
  \cellcolor{gray!50}\textbf{Average clock drift  (ms)} &
  \cellcolor{gray!50}\textbf{$\sigma$ of clock drift (ms)}&
  \cellcolor{gray!50}\textbf{Average introduced delay (ms)} &
  \cellcolor{gray!50}\textbf{$\sigma$ of introduced delay (ms)}\\
  \hline
$2$ & $-993$ & $824$ & $813$ & $2514$ \\ 
  $5$ & $-979$ & $832$ & $1173$ & $2446$ \\ 
  $6$ & $-938$ & $862$ & $702$ & $2326$ \\ 
  $7$ & $-850$ & $787$ & $896$ & $2281$ \\ 
   \hline
\end{tabular}
\caption{Delays observed in the experiments}
\label{tab:delay}
\end{table}

\subsection{Dual Prediction Schemes}

Experiments using real sensor data showed that state-of-the-art forecasting 
methods can be successfully implemented in the sensor nodes to keep the quality 
of their measurements and reduce up to $30\%$ of their 
transmissions~\cite{Dias2016b}.
Since WARM supports new applications, a data analytics algorithm can be used to 
compute parameters of forecasting methods (such as the AutoRegressive Integrate 
Moving Average--ARIMA).
Once these parameters are transmitted by the \dashboard{}, sensor nodes can 
predict measurements using simple arithmetic functions, which are not 
computing-intensive.
If a measurement differs by less than a certain threshold from its prediction, 
the sensor node will save a transmission, because the same prediction can be 
simultaneously computed by the data analytics algorithm.

\subsection{Sensing as a Service}

If a user management layer is attached to the \dashboard{} back-end, it will 
facilitate the identified communication with other systems and allow the 
information sharing, creating a new business model that offers the information 
retrieved by sensor nodes as a service.
For instance, WSNs can establish data-sharing agreements involving monetary
compensations for eventual cooperation and use shared data from one or more WSNs 
to optimize another WSN's performance, after analyzing and evaluating how the 
environment is changing at a certain moment.
Alternatively, users may pay for temperature measurements in a certain region.
Meanwhile, data analytics algorithms can improve the resource utilization and 
offer confidence intervals to estimated values locally computed.

\section{Conclusion}

In this work, we described the implementation of an architecture that 
integrates WSNs in IoT environments.
In practice, data gathered by WSNs can be displayed to their managers and other
stakeholders, such as data consumers and third-party services that can
benefit from the knowledge generated by sensor nodes. 
The differences from ``traditional'' WSNs are the online data analysis and the
capacity of self-management, thanks for the interconnection of managers 
at different planes.
The success of our deployment shows that WSNs can be incorporated into 
self-managing IoT environments that do not
depend upon human intervention for fine-tuning their operation.

To control WSNs, a specialized resource manager (WARM) was adopted.
It relies on SDN features to control and manage sensor nodes, avoiding
misuse and underutilization of WSNs' resources.
Connected to WARM, a WSN dashboard (\dashboard{}) facilitates the integration 
between WSNs, users and a \dataServer{}.
The \dashboard{} collects, stores and publishes data transmitted by wireless 
sensor nodes to the \dataServer{}, which can perform data analysis 
remotely in the cloud.

In the experiments, a real WSN was automatically configured using a 
reinforcement learning algorithm that learns from the historical data and 
generates instructions to adapt sensor nodes according to the environmental 
changes.
Results showed no human intervention was necessary to adjust the sampling 
intervals according to the environment's evolution and reduce up to 
$84.62\%$ the number of transmissions, which are the major cause of 
energy consumption in wireless sensor nodes.
Moreover, it was observed a small delay (less than $1.2$ seconds) introduced 
by the new data flow, which illustrates the architecture's real-timeliness that 
can be exploited in several use cases.

Thanks to the scalability of the proposed architecture, future works may 
incorporate several WSNs reporting data simultaneously to a \dashboard{} in the 
cloud and develop novel applications that exploit the measured data at the 
maximum.
Meanwhile, users will be able to visualize information about WSNs and monitor 
their data, according to their privileges in a user management system that 
may be attached to the \dashboard{}, which opens the possibility for a new 
business model to be explored.

\section*{Acknowledgment}

This paper is part of a project that has received funding from 
the European Union's 7th Framework Program under Grant 
Agreement no. 605073.
Also, this work has been partially supported by the Catalan Government 
through the project SGR-2014-1173.
C. B. Margi is supported by CNPq research fellowship \#307304/2015-9.

\bibliographystyle{IEEEtran}
\bibliography{IEEEabrv,bibliography}

\begin{thebibliography}{10}
\providecommand{\url}[1]{#1}
\csname url@samestyle\endcsname
\providecommand{\newblock}{\relax}
\providecommand{\bibinfo}[2]{#2}
\providecommand{\BIBentrySTDinterwordspacing}{\spaceskip=0pt\relax}
\providecommand{\BIBentryALTinterwordstretchfactor}{4}
\providecommand{\BIBentryALTinterwordspacing}{\spaceskip=\fontdimen2\font plus
\BIBentryALTinterwordstretchfactor\fontdimen3\font minus
  \fontdimen4\font\relax}
\providecommand{\BIBforeignlanguage}[2]{{%
\expandafter\ifx\csname l@#1\endcsname\relax
\typeout{** WARNING: IEEEtran.bst: No hyphenation pattern has been}%
\typeout{** loaded for the language `#1'. Using the pattern for}%
\typeout{** the default language instead.}%
\else
\language=\csname l@#1\endcsname
\fi
#2}}
\providecommand{\BIBdecl}{\relax}
\BIBdecl

\bibitem{Bellavista2013}
\BIBentryALTinterwordspacing
P.~Bellavista, G.~Cardone, A.~Corradi, and L.~Foschini, ``{Convergence of MANET
  and WSN in IoT Urban Scenarios},'' \emph{IEEE Sensors Journal}, vol.~13,
  no.~10, pp. 3558--3567, Oct. 2013. [Online]. Available:
  \url{http://ieeexplore.ieee.org/lpdocs/epic03/wrapper.htm?arnumber=6552998}
\BIBentrySTDinterwordspacing

\bibitem{Chatzigiannakis2010}
\BIBentryALTinterwordspacing
I.~Chatzigiannakis, S.~Fischer, C.~Koninis, G.~Mylonas, and D.~Pfisterer,
  \emph{Sensor Applications, Experimentation, and Logistics: First
  International Conference, SENSAPPEAL 2009, Athens, Greece, September 25,
  2009, Revised Selected Papers}.\hskip 1em plus 0.5em minus 0.4em\relax
  Berlin, Heidelberg: Springer Berlin Heidelberg, 2010, ch. WISEBED: An Open
  Large-Scale Wireless Sensor Network Testbed, pp. 68--87. [Online]. Available:
  \url{http://dx.doi.org/10.1007/978-3-642-11870-8_6}
\BIBentrySTDinterwordspacing

\bibitem{Ezdiani2016}
S.~Ezdiani, I.~S. Acharyya, S.~Sivakumar, and A.~Al-Anbuky, ``{An IoT
  Environment for WSN Adaptive QoS},'' \emph{Proceedings - 2015 IEEE
  International Conference on Data Science and Data Intensive Systems; 8th IEEE
  International Conference Cyber, Physical and Social Computing; 11th IEEE
  International Conference on Green Computing and Communications and 8th IEEE
  International Conference on Internet of Things, DSDIS/CPSCom/GreenCom/iThings
  2015}, pp. 586--593, 2016.

\bibitem{Madden:2005:TAQ:1061318.1061322}
\BIBentryALTinterwordspacing
S.~R. Madden, M.~J. Franklin, J.~M. Hellerstein, and W.~Hong, ``Tinydb: An
  acquisitional query processing system for sensor networks,'' \emph{ACM Trans.
  Database Syst.}, vol.~30, no.~1, pp. 122--173, Mar. 2005. [Online].
  Available: \url{http://doi.acm.org/10.1145/1061318.1061322}
\BIBentrySTDinterwordspacing

\bibitem{Minh2013}
\BIBentryALTinterwordspacing
T.~Minh, B.~Bellalta, and M.~Oliver, ``{DISON: A Self-organizing Network
  Management Framework for Wireless Sensor Networks},'' \emph{Ad Hoc Networks},
  2013. [Online]. Available:
  \url{http://link.springer.com/chapter/10.1007/978-3-642-36958-2\_11}
\BIBentrySTDinterwordspacing

\bibitem{Silv1608:WARM}
H.~Silva, A.~{Hahn Pereira}, Y.~Solano, B.~T. de~Oliveira, and C.~B. Margi,
  ``{WARM:} {WSN} application development and resource management,'' in
  \emph{XXXIV Simposio Brasileiro de Telecomunicacoes e Processamento de
  Sinais(SBrT 2016) (SBrT 2016)}, Santar{\'e}m, Brazil, Aug. 2016.

\bibitem{DeOliveira2014}
B.~T. {De Oliveira}, C.~B. Margi, and L.~B. Gabriel, ``{TinySDN: Enabling
  multiple controllers for software-defined wireless sensor networks},'' in
  \emph{2014 IEEE Latin-America Conference on Communications, IEEE LATINCOM
  2014}, 2014.

\bibitem{Dias2016c}
\BIBentryALTinterwordspacing
G.~M. Dias, T.~Adame, B.~Bellalta, and S.~Oechsner, ``{A self-managed
  architecture for sensor networks based on real time data analysis},'' to
  appear in Future Technologies Conference 2016.\hskip 1em plus 0.5em minus
  0.4em\relax IEEE, Dec. 2016. [Online]. Available:
  \url{https://arxiv.org/abs/1605.09011}
\BIBentrySTDinterwordspacing

\bibitem{riverbedmodeler2016}
\BIBentryALTinterwordspacing
I.~Riverbed~Technology. (2016, jun) Riverbed modeler. [Online]. Available:
  \url{http://www.riverbed.com/products/steelcentral/steelcentral-riverbed-modeler.html}
\BIBentrySTDinterwordspacing

\bibitem{Watkins1992}
\BIBentryALTinterwordspacing
C.~J. Watkins and P.~Dayan, ``Technical note: Q-learning,'' \emph{Machine
  Learning}, vol.~8, no.~3, pp. 279--292, 1992. [Online]. Available:
  \url{http://dx.doi.org/10.1023/A:1022676722315}
\BIBentrySTDinterwordspacing

\bibitem{Dias2016d}
\BIBentryALTinterwordspacing
G.~M. Dias, M.~Nurchis, B.~Bellalta, and S.~Oechsner, ``{Adapting sampling
  interval of sensors using reinforcement learning},'' in \emph{Submitted to
  the IEEE World Forum on Internet of Things 2016}.\hskip 1em plus 0.5em minus
  0.4em\relax IEEE, May 2016. [Online]. Available:
  \url{https://arxiv.org/abs/1606.02193}
\BIBentrySTDinterwordspacing

\bibitem{Dias2016b}
\BIBentryALTinterwordspacing
G.~M. Dias, B.~Bellalta, and S.~Oechsner, ``On the importance and feasibility
  of forecasting data in sensors,'' in \emph{Submitted to Transactions on
  Mobile Computing Journal}, 2016. [Online]. Available:
  \url{https://arxiv.org/abs/1604.01275}
\BIBentrySTDinterwordspacing

\end{thebibliography}

\end{document}